\documentstyle[aaspp4]{article}

\begin{document}

\def\sec{$^{\prime\prime}$}
\def\min{$^{\prime}$}

\title{The Difference Between the Narrow Line Region of Seyfert 1 and Seyfert 2 Galaxies}

\author{Henrique R. Schmitt \altaffilmark{1,2,3}}

\altaffiltext{1} {Depto. Astronomia, IF-UFRGS, CP 15051, CEP 91501-970, Porto Alegre, RS, Brazil}

\altaffiltext{2} {CNPq Fellow}

\altaffiltext{3} {email:schmitt@if.ufrgs.br}

\begin{abstract}

This paper presents a comparative study of emission line ratios of the
Narrow Line Region (NLR) of Seyfert 1 and Seyfert 2 galaxies. It includes
a literature compilation of the emission line
fluxes [OII]$\lambda$3727\AA, [NeIII]$\lambda$3869\AA,
[OIII]$\lambda$5007\AA\ and [NeV]$\lambda$3426\AA, as well as 60$\mu$m
continuum flux, for a sample of 52 Seyfert 1's and 68 Seyfert 2's.
The distribution of the emission line ratios [OII]/[NeIII]
and [OII]/[NeV] shows that Seyfert 1's and Seyfert 2's are
statistically different, in the sense that Seyfert 1's have values
smaller than those of Seyfert 2's, indicating a higher excitation
spectrum. These and other emission line ratios are compared with
sequences of models which combine different proportions of matter and
ionization bounded clouds and also sequences of models which vary only
the ionization parameter. This comparison shows that the former models
reproduce better the overall distribution of emission line ratios,
indicating that Seyfert 1's have a smaller number of ionization
bounded clouds than Seyfert 2's. This difference, together with other results
available in the literature, are interpreted from the point of view of four
different scenarios. The most likely scenario assumes that Seyfert 1's
have NLR's smaller than those of Seyfert 2's, possibly due to a
preferential alignment of the torus axis close to the host galaxy plane
axis in Seyfert 1's.

\end{abstract}


\section{Introduction}

The observation of broad polarized lines in the spectrum of the Seyfert 2
galaxy NGC1068 (Antonucci \& Miller 1985) showed that Seyfert 2's can be
Seyfert 1's where the direct view of the central engine is blocked.
This is the basis for the Unified Model of AGN's, which assumes that objects
of different activity class, like Seyfert 1's and Seyfert 2's, are the same
kind of object, surrounded by a dusty molecular torus. The orientation of
this torus relative to the line of sight determines whether the AGN is observed
as a broad lined object (Seyfert 1), where the nuclear engine is seen
through the torus opening, or as a narrow lined object (Seyfert 2),
where our view of the central engine and consequently the broad lines,
is blocked by the torus.

Several pieces of observational evidence supporting this
scenario have been obtained during the last decade, the strongest one being the
observation of polarized broad emission lines in the spectrum of several
Seyfert 2 galaxies (Antonucci \& Miller 1985; Miller \& Goodrich 1990;
Kay 1994; Tran 1995). The observation of collimated radiation escaping
the nuclear region of Seyfert 2's, seen as cone like emission line regions
(Pogge 1988a,b, 1989; Schmitt, Storchi-Bergmann \& Baldwin 1994,
Schmitt \& Storchi-Bergmann 1996, and references therein), or linear radio
structures (Ulvestad \& Wilson 1984a,b, 1989), also suggest that the direct
view of the central engine is blocked in these objects.
More direct evidence for the obscuration of the central engine in
Seyfert 2's comes from the analysis of X-ray spectra, which show large
absorbing column densities in these objects (Mulchaey,
Mushotzky \& Weaver 1992). Also, the observation of H$_2$O masers very
close to the nucleus of some Seyfert 2's, like NGC1068 and NGC4258
(Miyoshi 1995, Gallimore 1996, Greenhill 1996), show the presence of
large concentrations of molecular gas, hiding the central engine.

Recent papers, however, present some results suggesting that not only
the orientation of the circumnuclear torus relative to the line of
sight, but also its orientation relative to the host galaxy may be
important in the AGN classification. It was known since Keel (1980),
that there is a paucity of Seyfert 1's with edge-on host galaxies. This
result was later confirmed by Maiolino \& Rieke (1995) and Simcoe et al.
(1997), who suggested
that, in some cases, dust along a Seyfert 1 galaxy disk may be
responsible for the obscuration of the broad lines (making it appear
as a Seyfert 2). Moreover, Schmitt et al. (1997) presented a
comparison between the linear radio structure of Seyfert galaxies, with
their host galaxy major axis. They found that the radio structures are
more likely to be aligned close to the host galaxy plane axis in Seyfert
1's, but can have any direction in Seyfert 2's, confirming the result by
Maiolino \& Rieke (1995). Another result that corroborates this scenario is
the observation that the NLR of Seyfert 1's usually is much smaller
than that of Seyfert 2's, when they are compared as if Seyfert 2's were
observed pole-on, in the same way as Seyfert 1's (Schmitt \& Kinney
1996). The smaller Seyfert 1 NLR's can be understood if these objects
have their torus axis preferentialy aligned close to the host galaxy
plane axis, where there is less gas to be ionized.

The above results show differences between the NLR of Seyfert 1 and
Seyfert 2 galaxies and point towards older papers, where some other
differences have also been detected. Heckman \& Balick (1979) and
Shuder \& Osterbrock (1981) showed that the ratio [OIII]4363/5007
is larger in Seyfert 1's than in Seyfert 2's. This result indicates
that the [OIII] zone of Seyfert 1's, when compared to Seyfert 2's,
have larger temperatures and/or densities.
Yee(1980) and Shuder (1981) showed that
the emission lines, [OIII], [OII] and [OI], are more luminous in
Seyfert 2's relative to Seyfert 1's of similar optical luminosity,
consistent with the torus blocking part of the continuum light in
Seyfert 2's. Shuder \& Osterbrock (1981) and Cohen (1983) showed that the
emission line ratios [FeVII]/H$\beta$ and [FeX]/H$\beta$ are larger in
Seyfert 1's than in Seyfert 2's, indicating that Seyfert 1's have
higher excitation. Yet another interesting result was
obtained by De Robertis \& Osterbrock (1986 and references therein),
who showed that the FWHM of forbidden lines are well correlated with the
ionization potential in Seyfert 1's, but not with the critical  density
for de-excitation, while in Seyfert 2's the opposite happens. They have
also showed that these lines have smaller FWHM in Seyfert 1's than in
Seyfert 2's, and that the [OI] line profiles show evidence of two
components in Seyfert 2's, probably formed in two different regions.

This paper presents a compilation of literature data of the emission
line fluxes [OII]$\lambda$3727\AA, [NeIII]$\lambda$3869\AA,
[NeV]$\lambda$3426\AA\ and [OIII]$\lambda$5007\AA\ ([OII], [NeIII],
[NeV] and [OIII], hereafter), as well as 60$\mu$m continuum fluxes for
a sample of 52 Seyfert 1 and 68 Seyfert 2 galaxies. These lines are
used to compare the excitation of the NLR gas in Seyfert 1's and
Seyfert 2's, through the analysis of different emission line ratios.  A
simple interpretation of the Unified Scheme would suggest that the
spectrum of the NLR of Seyfert 1's and Seyfert 2's should have similar
degrees of excitation. However, as shown by the above papers, this may
not be true. Effects like the possible obscuration of parts of the NLR
by the torus, or the smaller NLR size in Seyfert 1's, could influence
the average NLR excitation in these two classes of objects.

The paper is organized in the following way, Section 2 presents the
sample, the reasons for the choice of these emission lines and a discussion
of the possible selection effects. Section 3 shows the results of the
comparison between Seyfert 1's and Seyfert 2's. Section 4 shows the
comparison between the data and photoionization models and discusses
possible interpretations of the results, while Section 5 gives the
summary.

\section{The Data and Selection Effects}

The usual way to analyze the gas excitation in galaxies is through the
use of ratios between different emission line fluxes. The most common
approach is to use BPT diagrams, which allow the
differentiation between Seyfert 2's, LINER's and HII regions (Baldwin,
Phillips \& Terlevich 1981). These diagrams use emission line ratios
like [OII]/[OIII], [NII]/H$\alpha$ and [OIII]/H$\beta$, which can be
easily measured in Seyfert 2 galaxies. However, due to blending with
broad lines, as is the case for H$\beta$ and H$\alpha+$[NII], these lines
cannot be easily measured in Seyfert 1's. Another problem
is the difficulty in determining the internal reddening in Seyfert 1's,
which can considerably influence the [OII]/[OIII] ratio.

In order to avoid the above problems, a different set of emission
lines, easily measurable in both Seyfert 1 and Seyfert 2 galaxies,
is chosen. These lines are: [OII]$\lambda$3727\AA, [NeIII]$\lambda$3869\AA,
[NeV]$\lambda$3426\AA\ and [OIII]$\lambda$5007\AA.  They span a
wide range in ionization potentials, are not blended with other lines,
either broad or narrow and, with the exception of [OIII], are close
in wavelength, which minimizes scatter resulting from reddening effects
and relative flux calibration errors. [NeIII] is of particular interest,
because its values of ionization potential and critical density
for collisional de-excitation are
very similar to those of [OIII], implying that they may be formed in
similar regions. In this way the ratio [OII]/[NeIII] can be used in the
place of  [OII]/[OIII] with the advantage of being reddening free.
For more on the use of [NeIII] in diagnostic diagrams, see
Rola, Terlevich \& Terlevich (1997).

The literature was searched for Seyfert 1 and Seyfert 2 galaxies with
measured emission line fluxes of the lines [OII], [NeIII], [NeV] and
[OIII].  It was possible to find 52 Seyfert 1 and 68 Seyfert 2
galaxies, which are shown in Tables 1 and 2, respectively. These Tables
give the names of the objects, B magnitude, radial velocity, Morphological
Type (de Vaucouleurs et al. 1991, Mulchaey 1994), the emission line ratios [OII]/[NeIII], [OII]/[NeV],
[NeIII]/[NeV] and [OII]/[OIII], the IRAS 60$\mu$m flux, the reference
from which the emission line ratios were obtained and the aperture size
used to observe the spectrum.  Notice that it was not possible to find
[NeV] and 60$\mu$m flux, as well as Morphological Type for all galaxies
in the sample.

An important point about the data collection is that, for every object,
emission line fluxes from two different references were never mixed.
Also, preference was given to data obtained with medium size aperture
(3\arcsec--7\arcsec), in order to include as much NLR emission as possible,
but also avoid extremely large aperture sizes, which could include HII regions
in the galaxy disk. The apertures given in
Tables 1 and 2 were classified in 3 categories, S (small),
corresponding to apertures smaller than 3\arcsec, M (medium),
corresponding to apertures in the range 3\arcsec--7\arcsec\ and L
(large), corresponding to apertures larger than 7\arcsec.

The radial velocities and aperture sizes (in arcseconds) were used to
calculate the metric aperture sizes, which correspond to the dimension
of the aperture in the galaxy (in parsecs), calculated assuming
H$_0=75$ km s$^{-1}$ Mpc$^{-1}$. These values were
calculated by taking the square root of the slit area and, when comparing
the average metric aperture sizes for Seyfert 1's and Seyfert 2's,
it was found that they have similar values, with means and 1$\sigma$
uncertanties of 1972$\pm$1578~pc and 2008$\pm$2101~pc, respectively.
The Spearman rank test was used to
compare the four emission line ratios with the metric aperture sizes.
They do not show any correlation, confirming that aperture
effects are not a problem for the analysis.

Since the sample was obtained from the literature, rather than selected
from an isotropic property, it is necessary to check if both Seyfert 1
and Seyfert 2 galaxies have similar intrinsic properties.  First it was
checked if Seyfert 1's and Seyfert 2's have similar luminosities and
are not biased towards high luminosity Seyfert 1's and low luminosity
Seyfert 2's, which could imply a larger flux of high excitation
lines in Seyfert 1's. This test was done by comparing the 60$\mu$m
luminosities of the two groups of galaxies.
Here it was assumed that the 60$\mu$m luminosity is nuclear radiation
absorbed by the circumnuclear torus and reradiated in the far-infrared,
so it should scale with total luminosity. However, notice that it can
depend on the torus covering factor, which can differ for Seyfert 1's
and Seyfert 2's. Also, it should be noticed that
the assumption that 60$\mu$m luminosity scale with the
nuclear luminosity must be taken with care, because, as pointed out
by Pier \& Krolik (1992), the torus emission may be anisotropic even at
60$\mu$m.

The results for this comparison are shown in Figure 1, where it can be
seen that both groups have similar distributions, with the
Kolmogorov-Smirnov (KS) test showing that two samples drawn from the
same parent population would differ this much 44\% of the time.  Table
3 gives the number of Seyfert 1 and Seyfert 2 galaxies with 60$\mu$m
luminosity available, their mean values, standard deviations, and the
KS test probability. This Table also gives information about other results from
the KS tests done below. The four emission line ratios were also compared with
the 60$\mu$m luminosity, but the Spearman rank test did not show any 
correlation. It should be noticed, however, that not all galaxies have IRAS 
60$\mu$m flux available, only 40 out of 52 Seyfert 1's and 52 out of 68 Seyfert
2's.

The second test checks whether the two groups have similar
morphological types and are not biased towards Seyfert 2's in late type
galaxies.  Late type galaxies are more likely to have circumnuclear HII
regions, which usually have much stronger [OII] than [NeIII] fluxes,
and would make Seyfert 2's look like lower excitation objects.  Figure
2 shows the distribution of morphological types, where it can be seen
that both groups have similar distributions, with the KS test showing
that two samples drawn from the same parent population would differ
this much 99.86\% of the time, in other words, would be more alike only
0.14\% of the time.

\section{Results}

Figure 3 shows the histogram of [OII]/[NeIII], where it can be seen
that Seyfert 1's have, on average, smaller values than those of Seyfert
2's, indicating a higher excitation spectrum. Of particular interest in
this histogram is the double cut-off in the distribution, for
[OII]/[NeIII]$<$1 there are 12 Seyfert 1's and only 2 Seyfert 2's, while for values
[OII]/[NeIII]$>$3.5 there are only Seyfert 2's.  Table 3 shows the
result of the KS test for this emission line ratio, which
shows that two samples drawn from the same parent population would differ this
much 0.02\% of the time.

Figure 4 shows the histogram of [OII]/[NeV]. As for [OII]/[NeIII],
Seyfert 1's again are displaced towards values smaller than
those found in Seyfert 2 galaxies. For [OII]/[NeV]$<$1 there are 17
Seyfert 1's and only 1 Seyfert 2. However, for this line ratio
there is not a high cut-off value, above which only Seyfert 2's are found, as
is the case for [OII]/[NeIII]. The KS test shows that two samples drawn
from the same parent population would differ this much only 0.16\% of the
time. It should be noticed that it was only possible to find [NeV]
fluxes for approximately 45\% of the galaxies in the sample.  This is
due in part to the fact that not all detectors have a good sensitivity
below $\lambda$3700\AA\ and to the fact that most of the NLR
studies are centered on emission lines above $\lambda$3700\AA. Also,
care must be taken when analyzing emission line ratios involving [NeV],
because the detection of this line can be biased towards high
excitation objects.

Figure 5 shows the distribution of [NeIII]/[NeV]. Besides the fact that
there are 7 Seyfert 1 and no Seyfert 2 galaxies with
[NeIII]/[NeV]$<$0.5, the two distributions are approximately similar,
with the KS test showing that two samples drawn from the same parent population
would differ this much only 11\% of the time.
However, as stated above, this result should be taken with
care, because it includes the [NeV] line.

The histogram of [OII]/[OIII] is shown in Figure 6, where it can be
seen that Seyfert 1's and Seyfert 2's have a similar distribution of
values. The KS test shows that two samples drawn from the same parent
population would differ this much only 23\% of the time.
This emission line ratio is shown here because it is one of the most common
indicators of gas excitation. However, due to the large wavelength
difference between the two lines ($\approx$1300\AA), this ratio is
extremely dependent on internal reddening and even the small value of
internal reddening E(B-V)=0.2 makes the [OII]/[OIII] ratio increase
by $\approx$20\%. As discussed above, [NeIII] originates in regions
similar to [OIII] and the [OII]/[NeIII] ratio can substitute
[OII]/[OIII].

\section{Discussion}

\subsection{Photoionization Models}

The results presented in the previous section show that the average
excitation of the NLR of Seyfert 1's is larger than that of Seyfert 2's.
This result is interpreted using diagnostic diagrams
involving the emission line ratios studied in this paper. Figures 7a,b,c
show the diagrams  Log~[OII]/[NeV]$\times$Log~[OII]/[NeIII],
Log~[NeIII]/[NeV]$\times$Log~[OII]/[NeIII] and
Log~[OII]/[OIII]$\times$Log~[OII]/[NeIII], respectively. It can be
seen that Seyfert 1's are more concentrated towards the lower left side
in these diagrams, which correspond to a higher excitation, confirming
the results obtained from Figures 3, 4, 5 and 6.
These distributions of emission line ratios are compared with
photoionization models, to analyze the possible origins of this difference
in excitation.

These results can be interpreted from the point of view of models that
combine different proportions of matter and ionization bounded clouds.
In these models the matter bounded clouds produce most of the high
excitation lines ([NeIII], [OIII] and [NeV]) and little of low
excitation lines ([OII] and [NII]), while the ionization bounded clouds
produce most of the low ionization lines and little of high excitation
lines.  The use of such models was proposed by Viegas \& Prieto (1992)
to explain the emission line region of 3C227. Later Binette, Wilson \&
Storchi-Bergmann (1996) (BWSB96, hereafter) used models of this kind to
study the extended NLR of Seyfert galaxies, showing their efficacy in
the reproduction of  high excitation lines, like [NeV]$\lambda$3426\AA\
and HeII$\lambda$4686\AA, as well as the [OIII]
temperature, which has always been a problem for the traditional
photoionization models which use sequences of ionization parameter.

Sequences of models adding different proportions of matter
and ionization bounded clouds were calculated using the photoionization
code MAPPINGS (Binette et al. 1993a,b), following the description given in
BWSB96. The models were calculated using a power law ionizing spectrum of
the form F$_{\nu}\propto\nu^{\alpha}$, and two different values of $\alpha$
were tested, --1.3 and --1.5.  The matter bounded clouds are ionized by this
spectrum and the calculation stops when 40\% of the incident spectrum
is absorbed. The output, reprocessed spectrum from the matter bounded
clouds, is the one that ionizes the ionization bounded clouds. The
models also assume that the ionization bounded clouds leak some of
the input radiation, in order to avoid overproduction of low ionization
lines, like [OII] and [NII]. In the case of $\alpha=-1.3$, it is assumed
that the ionization bounded clouds allow 3\% of the ionizing radiation to
escape, while for $\alpha=-1.5$ this value is 10\%.

The models were calculated considering an isobaric prescription, where
the pressure is constant
within any matter bounded or any ionization bounded cloud.  The
ionization parameter adopted for the matter bounded spectrum was
U=0.04. Nevertheless, for the ionization bounded clouds of the A$_{M/I}$
sequence (see below), instead of
specifying the ionization parameter, their pressure was fixed at 20
times that of the matter bounded clouds, as done by BWSB96. The adopted
density was n=50 cm$^{-3}$ and the gas metal abundance was solar (Z=1).
It is also assumed that the gas is mixed with a small quantity of dust
$\mu=0.015$\footnotemark and the abundance of metals in the grains is
depleted from the gas. Notice that this is a very small amount of dust and,
according to Binette et al. (1996), higher amounts of dust produce only
minimal changes in the output spectrum of ionization bounded clouds
but can have larger effects on the matter bounded clouds. However,
the matter bounded clouds are not expected to have a large quantity of
dust, because it can be easily destroyed by the radiation field.
The only independent parameter in these
models is the ratio between the solid angle subtended by matter bounded
clouds, relative to the solid angle subtended by the ionization bounded
clouds (A$_{M/I}$). Larger values correspond to a larger contribution
from matter bounded clouds relative to ionization bounded clouds and
vice-versa. This parameter was varied in the range
0.01$\leq$A$_{M/I}\leq634$, in steps of 0.2 dex.  These models are
represented as a solid line in Figures 7a,b,c, with the value of $\alpha$
indicated beside the line.

\footnotetext{$\mu$ is the dust to gas ratio of the clouds, in units of
the solar neighborhood dust-to-gas ratio}

In order to test the effects of other physical and chemical conditions,
two other sequences of A$_{M/I}$ models were calculated. In the first one
$\alpha=-1.5$, with the same parameters as above, but for gas
with twice the solar metalicity (Z=2). In the second set of models
$\alpha=-1.5$ and Z=1, but the density is 500 cm$^{-3}$. These
models have the same range of A$_{M/I}$ as above, are shown as a
long dashed lines in Figure 7a,b,c and are identified as Z=2 and n=500
cm$^{-3}$, respectively.

Just for comparison with the above models, MAPPINGS was also used to
calculate traditional sequences of models, varying only the ionization
parameter.  The parameters of the models were, power law ionizing spectrum
with $\alpha=-1.3$, constant density n=50 cm$^{-3}$, metalicity Z=1 and dust
content $\mu=$0.015. As for the A$_{M/I}$ models, two other sequences,
one with n=500 cm$^{-3}$ and Z=1, and the other with
n=50 cm$^{-3}$ and Z=2, were also tried.
It was assumed that 3\% of the ionizing radiation escape from the
clouds, in order to avoid the overproduction of low ionization lines.
The ionization parameter was varied in the range
$-4\leq$Log~U$\leq-0.8$, in steps of 0.2 dex.
The three sequences of models are very similar, the only 
exception being the sequence of models with Z=2 in the diagram
Log([OII]/[OIII])$\times$Log([OII]/[NeIII]) (Figure 7c), which are very similar
to the A$_{M/I}$ sequence with Z=2. Due to this fact, only the sequence
with n=50 cm$^{-3}$ and Z=1 is presented as a dotted line in Figures 7a,b,c.

It can be seen in Figures 7a and b, that the A$_{M/I}$ sequences of
models cover very well the observed distribution of values. In the case
of Figure 7c, the diagram Log([OII]/[OIII])$\times$Log([OII]/[NeIII]),
these models have some problem to reproduce the observed distribution
of values. It would be necessary
to change other parameters, like the amount of dust in the models, the
pressure jump between the matter and the ionization bounded clouds, or
the amount of radiation which leaks from the ionization bounded clouds,
in order to better reproduce the observed distribution of values.  The
fact that Seyfert 1's have [OII]/[NeIII] and [OII]/[NeV] values smaller
than Seyfert 2's, can be interpreted as due to a smaller contribution
from ionization bounded clouds, relative to matter bounded clouds,
to the spectra of those objects.  This
comparison also shows that the A$_{M/I}$ models with $\alpha=-1.3$ are
not as good a representation for the observed values as the ones with $\alpha=-1.5$, because they produce too much large fluxes of the higher 
excitation lines, like [NeV].

The comparison with the traditional U sequence of models, shows that
they are a poor representation of the data points, even when varying
parameters like the gas abundance or density.  Only in Figure 7c, where
the A$_{M/I}$ sequence of models has some problems to represent the
observed distribution of points, these models could be a better
representation for the data. However, they require unconventionally
large ionization parameters (U$>$0.01).

\subsection{Possible interpretations}

Four possible interpretations for the above result are studied here.

{\it 1-) Part of the matter bounded clouds (which produce most of
[NeIII], [OIII] and [NeV]),
is hidden by the circumnuclear torus in Seyfert 2's}. A similar problem
was found by Jackson \& Browne (1990) in the comparison of Quasars with
Radio Galaxies.
They show that the [OIII] emission of Quasars is much stronger than
that of Radio Galaxies, proposing that part of the [OIII] emission is
obscured by the torus in the latter objects. Hes, Barthel \& Fosbury (1993)
showed that, when comparing the [OII] emission of Quasars and Radio
Galaxies, which comes from a less obscured, lower excitation
region, both classes of objects have very similar distributions,
corroborating the obscuration scenario.

While the obscuration scenario can be the solution for Radio Galaxies
and Quasars, it may not be the general case for Seyfert 2 galaxies.
Assuming that the [OII] emission in Seyfert 2's is similar to that of
Seyfert 1's and not blocked by the torus, we can calculate, using the
average values given in Table 3, that $\approx$40\% of the [NeIII] emission,
$\approx$55\% of the [NeV] emission and $\approx$25\% of the [OIII]
emission should be blocked by the torus in Seyfert 2's. This could
happen for some of the Seyfert 2's in the sample, but notice that these
are large values and go against the fact that
Seyfert 2's have lower excitation lines (like [OII]) more luminous
than Seyfert 1's of similar optical luminosity
(Yee 1980; Shuder 1981). Also, Seyfert 2's
usually have extended NLR's (Pogge 1989; Schmitt \& Kinney 1996).
Another fact that goes
against the obscuration scenario being the general case is that, if part of
the high excitation emission line region is hidden by the torus, we would
expect to see considerable amounts of polarized [OIII] emission in
Seyfert 2's. As shown by Goodrich (1992), with a small number of
exceptions, Seyfert 2's do not have high degrees of polarized [OIII]
emission.

{\it 2-) We see a smaller number of ionization bounded clouds in Seyfert
1's, because they are seen from the back and are extincted.} Since the
ionization bounded clouds are responsible for most of the [OII]
emission and very little of the [NeIII], [NeV] and [OIII], this would
imply a reduction of the ratios [OII]/[NeIII], [OII]/[NeV] and
[OII]/[OIII] in Seyfert 1's, relative to Seyfert 2's.

>From the analysis
of the X-ray spectra of Seyfert 1's (Reynolds 1996; Weaver, Arnaud \&
Mushotzky 1995), it is known that they usually have small column
densities of absorbing material (N$_{HI}<10^{21}$ cm$^{-2}$).
Assuming a standard dust-to-gas ratio (A$_V=5\times10^{-22}$N$_{HI}$),
it is possible to estimate a typical value of extinction from the above
N$_{HI}$, which is A$_V<0.5$ (E(B-V)$\approx$0.2). In the case of
E(B-V)=0.1, the [OII] emission of the ionization bounded clouds would
be reduced by $\approx$35\%, which could explain the difference between
Seyfert 1's and Seyfert 2's. However, this scenario only works
when the ionization bounded clouds do not block the direct view of
the matter bounded clouds, otherwise the high excitation lines
would also be obscured.

{\it 3-) There is a smaller number of ionization bounded clouds in Seyfert 1's,
possibly due to the orientation of the circumnuclear torus relative
to the galaxy plane.} In thisscenario Seyfert 1's have
their circumnuclear torus axis preferentially aligned closer to the
host galaxy plane axis, while in Seyfert 2's the torus can have any
orientation. In this way, Seyfert 1's would have smaller NLR's,
because their ionizing radiation would shine out of the galaxy disk and
find only a small number of clouds to be ionized, thus resulting in
a smaller number of ionization bounded clouds in these objects. On the
other hand, since the Seyfert 2's torus axis can have any orientation
relative to the host galaxy disk, there is a larger chance for the ionizing
radiation to cross the galaxy disk in this objects, which would result in a
larger quantity of gas clouds to be ionized.
The clouds closer to the nucleus filter the ionizing radiation
and the more distant clouds are ionized only by this fainter and filtered
continuum. Due to the larger number of clouds along the disk, the nuclear
radiation ionizes a larger number of clouds, and this effect is similar
to be seeing a larger number of ionization bounded clouds in Seyfert 2's.

Some of the results available in the literature, discussed in the
introduction, corroborate this scenario. Seyfert 1's have higher
[OIII]4363/5007 ratios than Seyfert 2's, which could be explained as
higher [OIII] temperatures, or higher densities. If the higher
[OIII]4363/5007 ratios of Seyfert 1's
are in fact due to a higher [OIII] temperature,
this is consistent with a smaller proportion of ionization bounded clouds
in these objects, as shown by Binette et al. (1996) models.
This interpretation can also explain the results obtained by Schmitt \&
Kinney (1996), that Seyfert 1's have much smaller NLR's than Seyfert
2's (when they are compared in a similar way, as if they were seen
pole-on). Kraemer et al. (1998) confirmed this to the individual case
of the Seyfert 1 galaxy NGC5548, showing that this galaxy 
have a compact NLR, with a size of the order of 70pc.

The above results imply that the NLR of Seyfert 1's have less gas than
the NLR of Seyfert 2's, which can be explained if the Seyfert 1's torus
axis is aligned closer to the host galaxy plane axis.
This scenario is supported by the observation of a lack of Seyfert 1's
in edge-on galaxies (Keel 1980; Maiolino \& Rieke 1995; Simcoe et al. 1997)
and by the relative orientation between linear radio  structures
and the host galaxy major axis in Seyfert 1's (Schmitt et al. 1997).

{\it 4-) Seyfert 2's are more associated with circumnuclear star formation
(high metallicity HII regions) than Seyfert 1's.} Since high metallicity
HII regions are strong emitters of [OII] and weak emitters of [NeIII],
if the nuclear emission of Seyfert 2's is more likely to be mixed with HII
regions than Seyfert 1's, this would explain the fact that their NLR's show
less excited gas. Some evidence for the existence of circumnuclear regions
in Seyfert 2's is given by Heckman et al. (1995), Heckman et al. (1997),
Thuan (1984). However, this evidence is restricted to a small number of
galaxies and it would be necessary to study the stellar population of
a complete sample of Seyfert 1's and Seyfert 2's, in order to see if there
is any difference between these two classes of objects and if Seyfert 2's
in fact have more circumnuclear star formation. One such attempt
was done by Schmitt, Storchi-Bergmann \& Cid Fernandes (1998), who
synthesized the nuclear stellar population of 20 Seyfert 2's, showing that
young stars usually contribute with less than 5\% (less than 1\% in more
than 50\% of the sample) to the light of these galaxies at $\lambda$5870\AA.

\section{Summary}

This paper follows from a literature search of the fluxes of the
emission lines [OII], [NeIII], [NeV], [OIII] and of the 60$\mu$m
continuum for a sample of 52 Seyfert 1 and 68 Seyfert 2 galaxies. The
analysis of possible selection effects shows that the two groups are not
biased with respect to morphological type, have similar values of 60$\mu$m
luminosity, and were observed with apertures of similar metric sizes.

The comparison between the distribution of the emission line ratios
[OII]/[NeIII] and [OII]/[NeV] in Seyfert 1's and Seyfert 2's,
shows that the two groups are considerably different,
with  the Seyfert 2's spectra presenting more low excitation emission
than the spectra of Seyfert 1's.  The emission line ratios are compared
with sequences of models in which only the ionization parameter varies, as
well as with models which combine different proportions of matter and
ionization bounded clouds. It is shown that the distribution of observed
points can be better represented by the latter models, with Seyfert 1's
having a smaller number of ionization bounded clouds than Seyfert 2's.

Four possible interpretations for this difference are proposed. The most
likely explanation is that Seyfert 1's have smaller NLR's, thus
have a smaller number of ionization bounded clouds.
The NLR's of Seyfert 1's could be smaller than
those of  Seyfert 2's, due to an inclination effect. There is a growing
amount of evidence, showing that the torus axis of Seyfert 1's is more
likely to be aligned close to the galaxy plane axis, while in Seyfert 2's
it can have any direction (Schmitt et al. 1997; Simcoe et al. 1997).
In this way, the amount of gas ionized by the
nuclear radiation would be smaller in Seyfert 1's than in Seyfert 2's,
resulting in a larger number of ionization bounded clouds in these objects.

Two possibilities assume that part of the matter bounded clouds
is hidden by the circumnuclear torus in Seyfert 2's, or that the 
ionization bounded clouds are seen from the back in Seyfert 1's, creating
the impression that Seyfert 1's are more excited than Seyfert 2's.
The evidence presented above go against these two scenarios as a general
case. However, it is not possible to rule out individual cases where
they could happen.

A fourth possibility assumes that Seyfert 2's have a larger number of
circumnuclear star forming regions, relative to Seyfert 1's. There is
some evidence of circumnuclear starformation in some Seyfert 2's.
However, it still should be determined if this happens for all Seyfert 2's
and if there is a difference between the stellar population of the two
types of Seyferts.

It should be noticed that the results presented in this paper were
obtained from a sample selected from the literature, rather than a
sample selected from an isotropic property. Although it was shown that
the two samples have similar intrinsic properties, there could still be
some selection effects affecting the results. In order to avoid this,
it would be important to test these results using homogeneous
measurements of a sample selected by an isotropic property.

\acknowledgements 
I would like to thank  R. Antonucci, L. Binette, A. Kinney,
T. Storchi-Bergmann, C. Winge and the anonymous referee
for useful comments and suggestions. 
L. Binette is also thanked for making available
the code MAPPINGS. This research has made use of the NASA/IPAC
Extragalactic Database (NED) which is operated by the Jet Propulsion
Lab, Caltech, under contract with NASA.

\clearpage

\begin{figure}
\caption{Comparison between the 60$\mu$m luminosity of Seyfert 1's (dotted line)
and Seyfert 2's (solid line).}
\end{figure}

\begin{figure}
\caption{Comparison between the Morphological Types of the host galaxies of
Seyfert 1's (dotted line) and Seyfert 2's (solid line).}
\end{figure}

\begin{figure}
\caption{Comparison between the [OII]/[NeIII] distribution of Seyfert 1's
(dotted line) and Seyfert 2's (solid line).}
\end{figure}

\begin{figure}
\caption{Comparison between the [OII]/[NeV] distribution of Seyfert 1's
(dotted line) and Seyfert 2's (solid line).}
\end{figure}

\begin{figure}
\caption{Comparison between the [NeIII]/[NeV] distribution of Seyfert 1's
(dotted line) and Seyfert 2's (solid line).}
\end{figure}

\begin{figure}
\caption{Comparison between the [OII]/[OIII] distribution of Seyfert 1's
(dotted line) and Seyfert 2's (solid line).}
\end{figure}

\begin{figure} \caption{a-) Comparison between the emission line ratios
Log~[OII]/[NeV]$\times$Log~[OII]/[NeIII] and photoionization models.
Open symbols are Seyfert 1's and filled symbols are Seyfert 2's. The
solid lines represent the A$_{M/I}$ sequences of models with 
n=50 cm$^{-3}$, Z=1, ionized by a power law continuum
with slope $\alpha=-1.3$, or  $\alpha=-1.5$, as indicated
beside the line.  The dashed lines represent
the A$_{M/I}$ sequences of models ionized by power law spectra with
$\alpha=-1.5$, but n=500 cm$^{-3}$ and Z=1, or n=50 cm$^{-3}$ and Z=2,
as indicated beside the line by n=500 cm$^{-3}$ and Z=2, respectively.
A$_{M/I}$ was varied in the range 0.01$\leq$A$_{M/I}\leq$634 in steps
of 0.2 dex and decreases from left to right in the plot. The stars
along the lines are separated by 0.2 dex and the large star corresponds
to A$_{M/I}=4$.  Dotted line represents the sequence of ionization
parameter models, calculated using a power law
ionizing spectrum with $\alpha=-1.3$,
n=50 cm$^{-3}$ and Z=1. U was varied in the range
$-4\leq$Log~U$\leq-0.8$ in steps of 0.2 dex. The stars along the line
are separated by 0.2 dex, with the large star corresponding to
Log~U=--2.}
\end{figure}

\setcounter{figure}{6}

\begin{figure}
\caption{  b-)Same as Figure 7a, for the diagram
Log~[NeIII]/[NeV]$\times$Log~[OII]/[NeIII].}
\end{figure}

\setcounter{figure}{6}

\begin{figure}
\caption{  c-)Same as Figure 7a,
for the diagram Log~[OII]/[OIII]$\times$Log~[OII]/[NeIII]. Due to the fact that
the models with $\alpha=-1.3$ and $\alpha=-1.5$ are very similar, the plot
only shows the models with $\alpha=-1.5$.}
\end{figure}

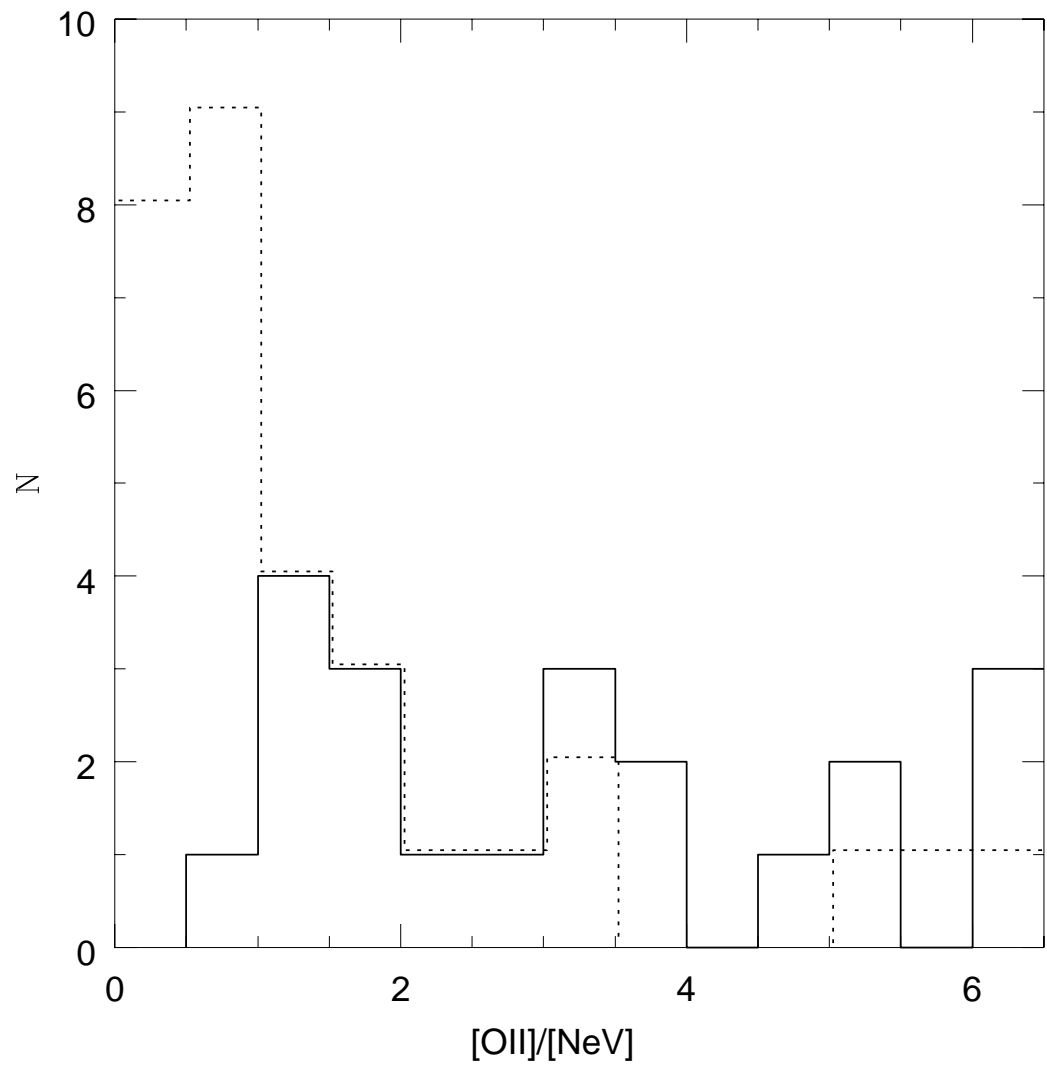
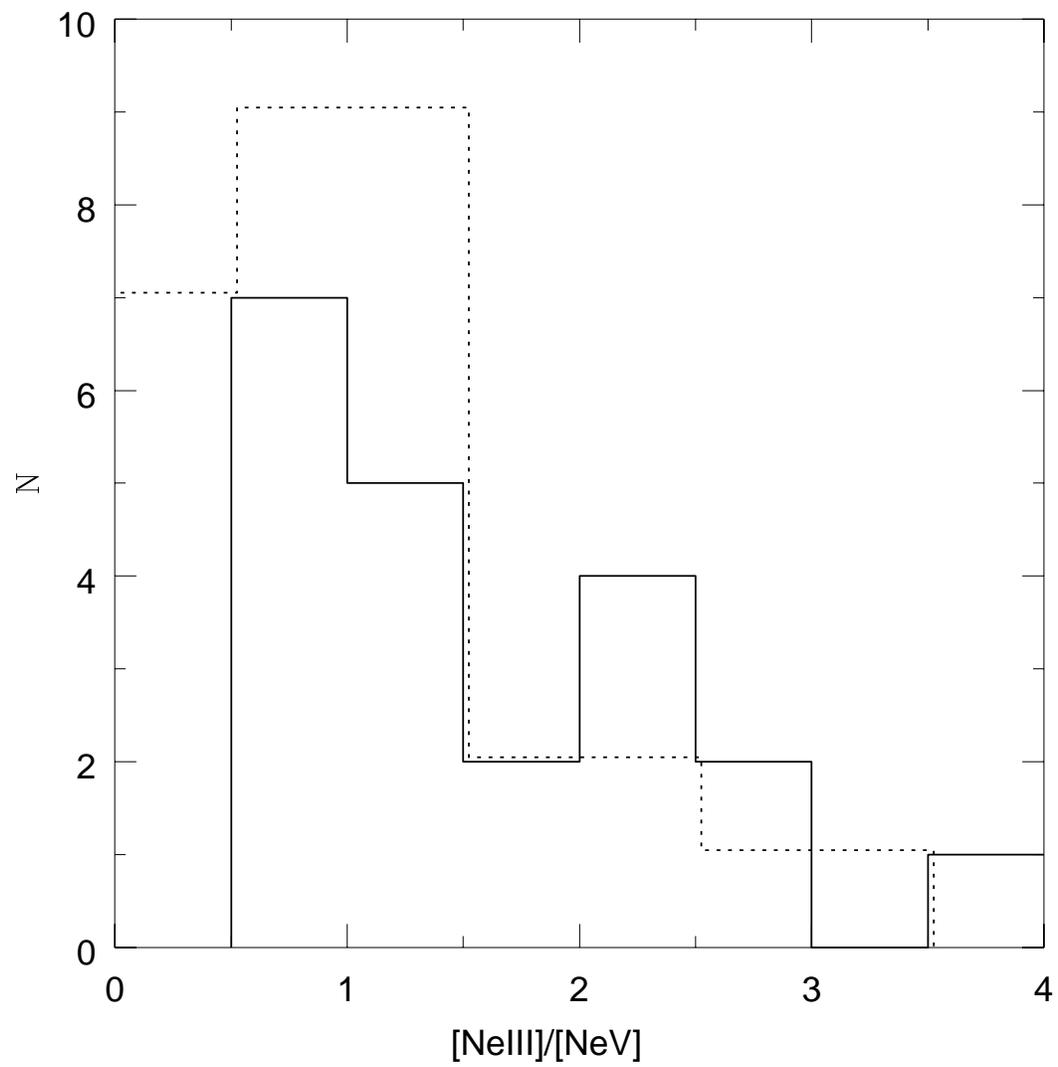
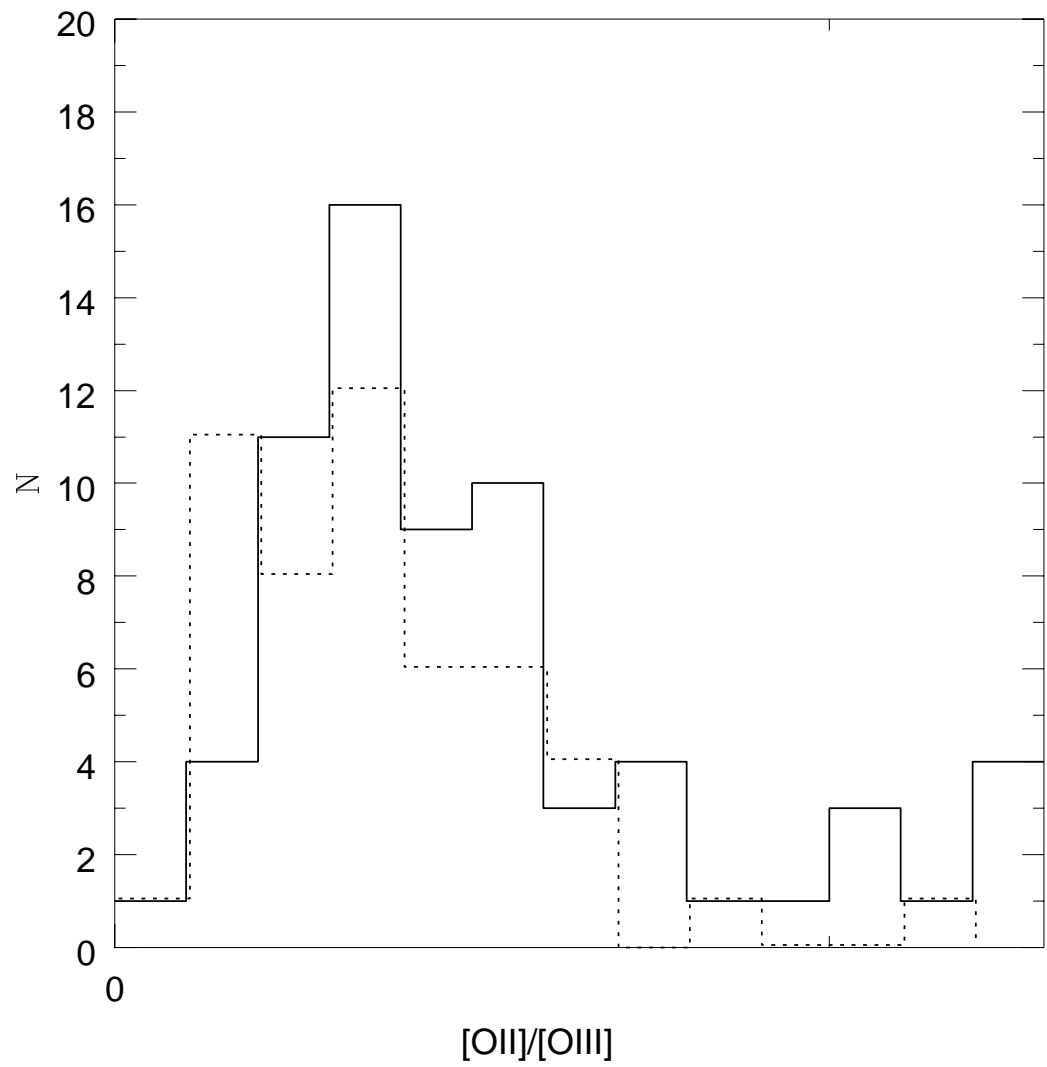
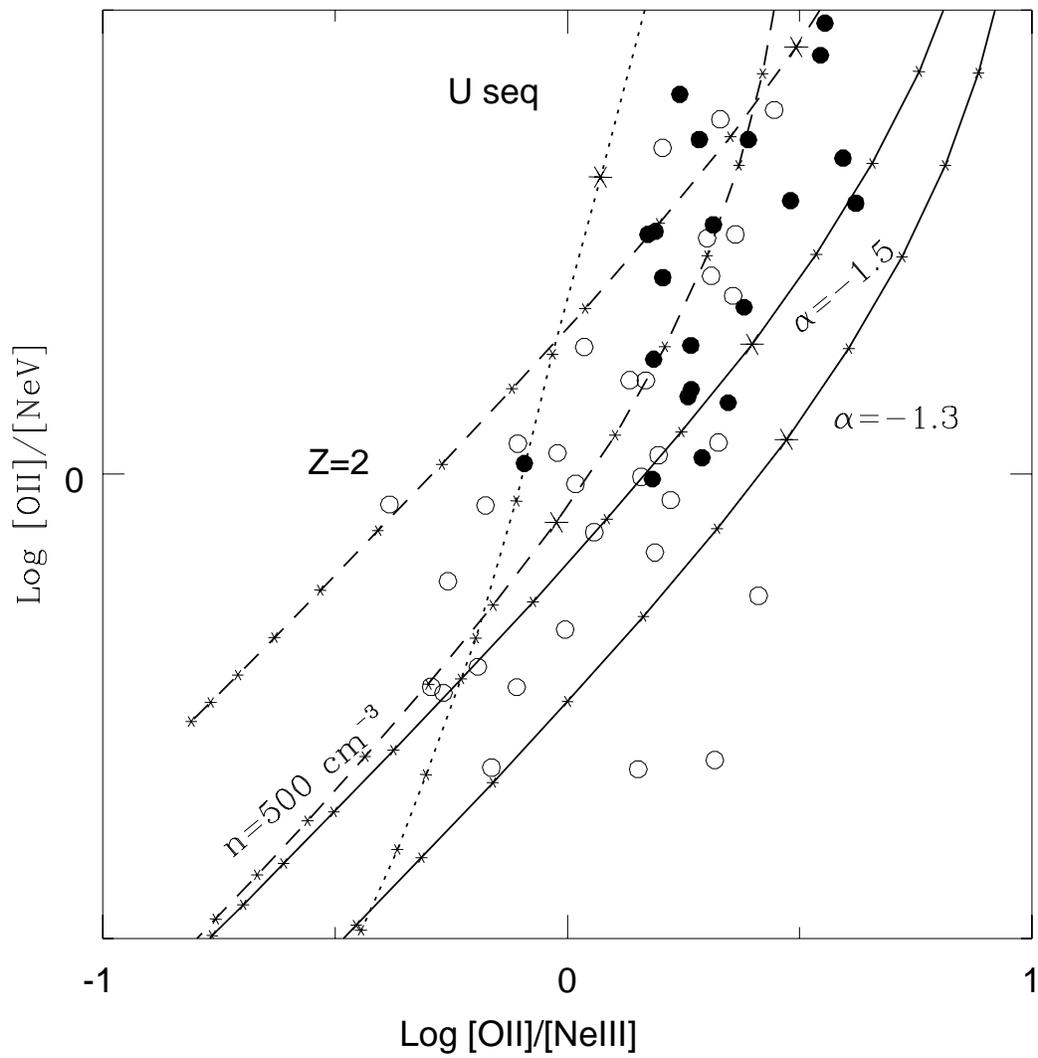
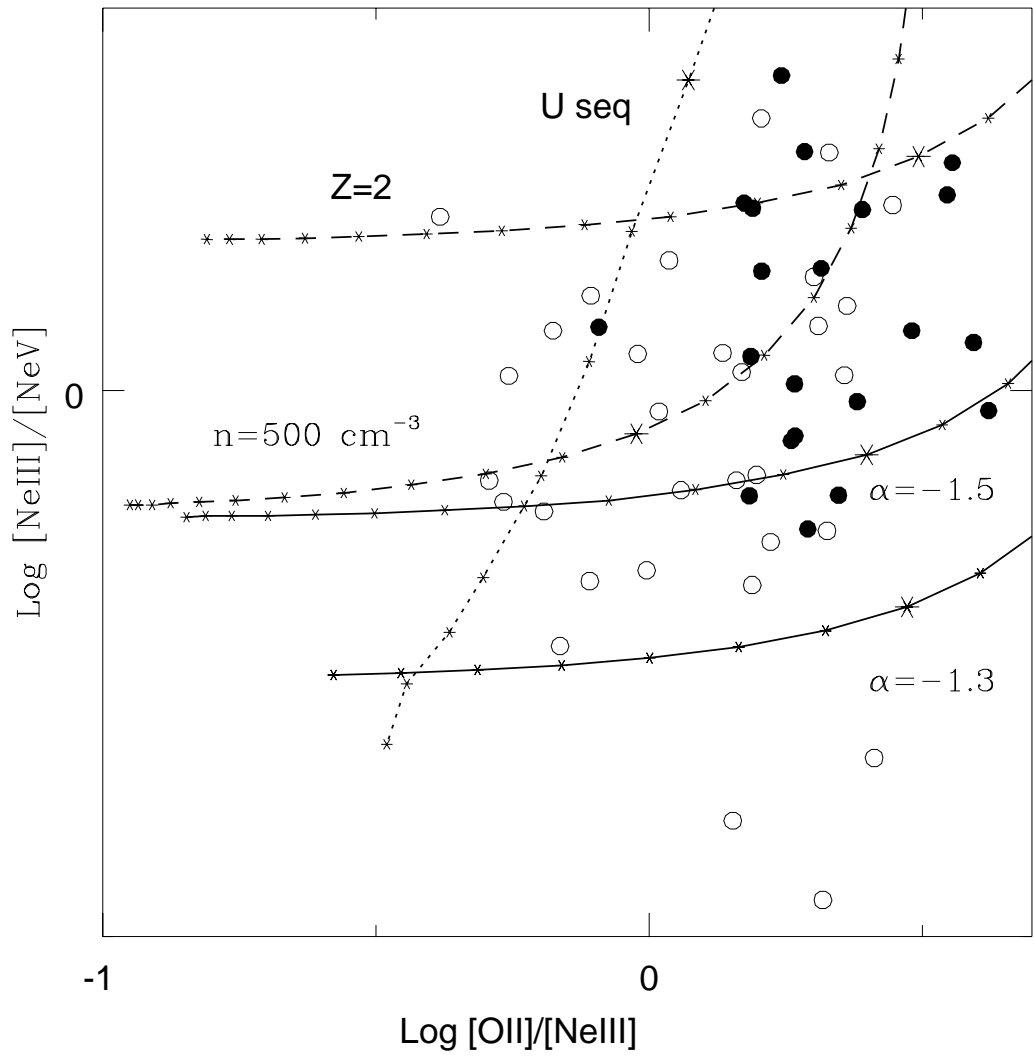
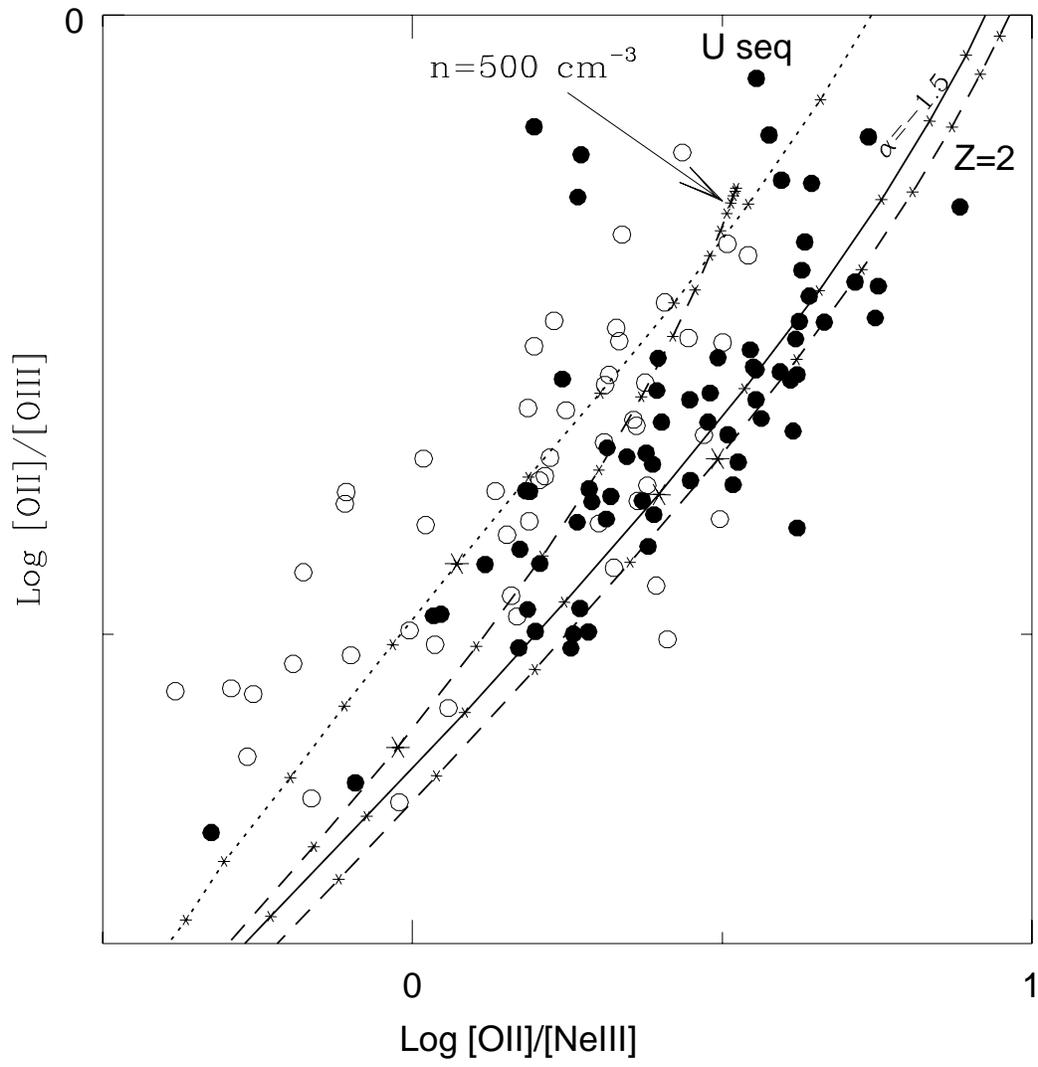
\end{document}




\small

\begin{deluxetable}{lrrlrrrrrrr}
\tablewidth{0pc}
\tablecaption{Seyfert 1's}
\tablehead{\colhead{Name}&\colhead{B}&\colhead{V$_0$}&\colhead{Morph.}&
\colhead{[OII]/[NeIII]}&\colhead{[OII]/[NeV]}&\colhead{[NeIII]/[NeV]}&\colhead{[OII]/[OIII]}&\colhead{F$_{60\mu m}$}&
\colhead{Ref.}&\colhead{Apert.}\cr
&&km s$^{-1}$&&&&&&(Jy)&&}
\startdata
NGC 1019	&14.95	&7290	&SBbc		&2.955	&---	&---	&0.210	&0.355	&12&M\cr
NGC 1566	&13.17	&1290	&SABbc		&2.375	&---	&---	&0.255	&14.71	&15&S\cr
NGC 3227	&13.52	&990	&SABa pc	&1.603	&5.05	&3.150	&0.177	&7.825	&22&S\cr
NGC 3516	&12.40	&2700	&SB0		&0.667	&0.857	&1.286	&0.126	&1.758	&16&L\cr
NGC 3783	&13.43	&2880	&SBa		&0.414	&0.861	&2.080	&0.081	&3.257	&14&M\cr
NGC 4051	&12.65	&600	&SABbc		&1.571	&1.100	&0.700	&0.292	&7.131	&16&L\cr
NGC 4151	&11.85	&900	&SABab		&1.087	&1.880	&1.729	&0.096	&---	&29&S\cr
NGC 4253	&13.60	&3810	&SB0/a		&0.952	&1.111	&1.167	&0.053	&4.026	&4&M\cr
NGC 4593	&13.15	&2610	&SBb		&0.782	&1.165	&1.490	&0.170	&3.052	&14&M\cr
NGC 5033	&10.75	&892	&SAc		&2.727	&---	&---	&0.600	&13.8	&26&S\cr
NGC 5548	&13.73	&5100	&SA0/a		&2.579	&0.547	&0.212	&0.098	&1.073	&22&S\cr
NGC 6814	&14.21	&1590	&SABbc		&1.443	&0.988	&0.685	&0.115	&5.517	&14&M\cr
NGC 6860	&13.50	&4470	&SBb		&2.556	&---	&---	&0.343	&0.954	&3&S\cr
NGC 7450	&14.33	&3120	&SBa		&2.115	&1.170	&0.553	&0.128	&---	&4&M\cr
NGC 7469	&13.15	&4830	&SABa		&2.310	&---	&---	&0.164	&25.87	&14&M\cr
Mrk 6		&14.19	&5610	&SAB0$^+$ 	&2.394	&---	&---	&0.174	&1.183	&7&M\cr
Mrk 42		&15.28	&7350	&		&2.157	&---	&---	&0.297	&0.317	&4&M\cr
Mrk 79		&14.02	&6570	&SBb		&1.362	&1.595	&1.171	&0.170	&1.503	&22&S\cr
Mrk 279		&14.46	&9150	&S0		&3.224	&---	&---	&0.427	&1.255	&22&S\cr
Mrk 315		&14.78	&11820	&S0/a pc	&2.179	&---	&---	&0.442	&1.464	&7&M\cr
Mrk 359		&14.22	&5072	&SB0		&1.143	&0.750	&0.656	&0.076	&1.132	&4&M\cr
Mrk 372		&14.81	&9300	&S0/a		&2.046	&---	&---	&0.253	&0.303	&7&M\cr
Mrk 506		&14.68	&12900	&SABa		&3.136	&---	&---	&0.153	&---	&22&S\cr
Mrk 509		&13.12	&10650	&E/S0		&2.077	&0.242	&0.117	&0.262	&1.364	&14&M\cr
Mrk 595		&14.69	&8250	&E/S0		&1.769	&---	&---	&0.230	&---	&17&M\cr
Mrk 704		&14.23	&8730	&Sa		&0.509	&0.348	&0.684	&0.082	&0.364	&22&S\cr
Mrk 783		&16.00	&20000	&		&3.167	&---	&---	&0.296	&0.31	&4&M\cr
Mrk 817		&13.90	&9600	&S0/a pc	&0.687	&0.234	&0.340	&0.054	&2.118	&22&S\cr
Mrk 841		&14.85	&10950	&		&1.042	&0.953	&0.915	&0.192	&0.459	&14&M\cr
Mrk 871		&14.80	&10200	&SB0		&0.989	&0.463	&0.468	&0.101	&0.69	&14&M\cr
Mrk 896		&14.61	&7860	&S?		&1.543	&0.679	&0.440	&0.152	&0.513	&14&M\cr
Mrk 926		&14.20	&14400	&Sa		&2.133	&5.818	&2.727	&0.312	&---	&23&S\cr
Mrk 975		&14.95	&14730	&S pec		&0.541	&0.338	&0.625	&0.063	&0.8	&22&S\cr
Mrk 1018	&14.30	&12810	&S0		&1.636	&---	&---	&0.180	&---	&25&S\cr
Mrk 1239	&14.39	&5820	&compact 	&1.050	&---	&---	&0.150	&1.335	&4&M\cr
Mrk 1320	&15.00	&30900	&SBb		&3.482	&---	&---	&0.409	&0.218	&14&M\cr
IC 4218		&14.40	&5820	&Sb-c		&1.536	&---	&---	&0.232	&---	&14&M\cr
IC 4329a	&13.66	&4800	&S0$^+$		&1.2	&---	&---	&0.028	&2.03	&28&S\cr
MCG 8-11-11	&14.00	&6150	&SBb-c		&2	&3.227	&1.614	&0.151	&3.005	&22&S\cr
MCG-6-30-15	&13.61	&2340	&E/S0		&2.273	&2.423	&1.066	&0.222	&1.087	&14&M\cr
UM 146		&14.50	&5160	&SAb		&2.475	&---	&---	&0.120	&0.467	&12&M\cr
Fairall 51	&14.10	&4080	&SBb		&1.668	&0.880	&0.528	&0.193	&1.844	&14&M\cr
ESO 141-G55	&13.60	&11040	&Sc		&0.795	&---	&---	&0.092	&0.575	&14&M\cr
UGC 10683b	&15.55	&9200	&		&1.421	&0.231	&0.163	&0.145	&0.479	&14&M\cr
TOL 0343-397	&14.83	&12900	&E		&1.692	&---	&---	&0.321	&0.24	&13&L\cr
TOL 1351-373	&	&15500	&		&1.476	&1.594	&1.080	&0.107	&0.45	&14&M\cr
TOL 1506.3-00	&	&16200	&		&2.039	&2.676	&1.312	&0.204	&---	&14&M\cr
TOL 20		&	&7000	&		&0.553	&0.588	&1.063	&0.080	&---	&14&M\cr
C 16.16		&17.19	&22864	&		&2.3	&3.286	&1.429	&0.217	&---	&5&S\cr
E 1615+061	&	&11370	&		&2.789	&6.092	&2.185	&0.301	&---	&14&M\cr
H 1839-78	&	&22200	&		&0.778	&0.348	&0.447	&0.162	&---	&14&M\cr
IIIZw 77	&	&10250	&		&0.642	&0.385	&0.600	&0.090	&0.245	&20&M\cr
\tablenotetext{}{Morphological types were obtained from 
de Vaucouleurs et al. 1991 and Mulchaey 1994. Column 9 gives the references
for the emission lines.
1-)Storchi-Bergmann, Bica, \& Pastoriza 1990; 2-)Storchi-Bergmann \& Pastoriza 1989;
3-)Lipari, Tsvetanov \& Macchetto 1993; 4-)Osterbrock \& Pogge 1985; 5-)Rodriguez-Ardila et al. 1996;
6-)Durret \& Bergeron 1986; 7-)Koski 1978; 8-)Osterbrock 1985; 9-)Diaz, Prieto \& Wamsteker 1988;
10-)Moran et al. 1992; 11-)Gonzalez Delgado \& Perez 1996; 12-)Phillips, Charles \& Baldwin 1983;	
13-)Terlevich et al. 1991; 14-)Morris \& Ward 1988; 15-)Alloin et al. 1985; 16-)Anderson 1970;
17-)Ulvestad \& Wilson 1983; 18-)Shuder \& Osterbrock 1981; 19-)Osterbrock \& Dahari 1983;
20-)Ferland \& Osterbrock 1986; 21-)Goodrich \& Osterbrock 1983; 22-)Cohen 1983;
23-)Durret \& Bergeron 1988; 24-)Kunth, Sargent \& Bothum 1987; 25-)Osterbrock 1981;
26-)Shuder 1980; 27-)Ward et al. 1980; 28-)Wilson \& Penston 1979; 29-)Boksenberg et al. 1975;
30-)Alloin et al. (1992).}
\enddata
\end{deluxetable}
\newpage
																							
\begin{deluxetable}{lrrlrrrrrrr}
\tablewidth{0pc}
\tablecaption{Seyfert 2's}
\tablehead{\colhead{Name}&\colhead{B}&\colhead{V$_0$}&\colhead{Morph.}&
\colhead{[OII]/[NeIII]}&\colhead{[OII]/[NeV]}&\colhead{[NeIII]/[NeV]}&\colhead{[OII]/[OIII]}&\colhead{F$_{60\mu m}$}&
\colhead{Ref.}&\colhead{Apert.}\cr
&&km s$^{-1}$&&&&&&(Jy)&&}
\startdata
NGC 424		&14.12	&3300	&SB0/a		&1.310	&---	&---	&0.130	&1.796	&22&S\cr
NGC 1068	&10.83	&1020	&Sb		&0.809	&1.056	&1.306	&0.058	&181.95	&7&M\cr
NGC 1229	&14.00	&10590	&SBb pec	&4.179	&3.836	&0.918	&0.263	&1.548	&12&M\cr
NGC 1358	&13.05	&3870	&SAB0/a		&4.253	&---	&---	&0.387	&0.378	&2&L\cr
NGC 1386	&12.84	&810	&Sa/S0		&2.810	&---	&---	&0.177	&5.396	&2&L\cr
NGC 1667	&12.77	&4472	&Sc		&5.182	&---	&---	&0.371	&5.952	&12&M\cr
NGC 2110	&13.51	&2130	&SA0		&4.300	&---	&---	&0.430	&4.129	&26&S\cr
NGC 2992	&13.78	&2250	&Sa pec		&4.623	&---	&---	&0.319	&6.760	&1&L\cr
NGC 3081	&13.55	&2160	&SAB0/a		&1.949	&1.086	&0.557	&0.164	&---	&6&S\cr
NGC 3281	&14.02	&3540	&SAa		&3.022	&3.886	&1.286	&0.245	&6.861	&23&S\cr
NGC 3393	&12.40	&3690	&Sa		&4.179	&---	&---	&0.148	&2.251	&9&M\cr
NGC 3786	&13.74	&2760	&SABa pc	&3.553	&---	&---	&0.270	&---	&21&S\cr
NGC 3982	&11.70	&10800	&SABb		&2.056	&---	&---	&0.153	&6.567	&12&M\cr
NGC 4074	&14.44	&6600	&S0 pec		&1.818	&1.471	&0.809	&0.100	&---	&18&S\cr
NGC 4388	&11.76	&2545	&SAb		&5.586	&---	&---	&0.324	&10.240	&1&L\cr
NGC 4507	&13.54	&3480	&SBab		&3.921	&4.798	&1.224	&0.265	&4.310	&6&S\cr
NGC 4941	&12.23	&870	&SABab		&4.079	&---	&---	&0.257	&1.378	&2&L\cr
NGC 4939	&13.80	&2910	&SAbc		&1.484	&---	&---	&0.095	&2.015	&1&L\cr
NGC 5135	&13.35	&3990	&SBab		&2.524	&---	&---	&0.220	&16.910	&12&M\cr
NGC 5256	&13.42	&8280	&S0 pec		&5.446	&---	&---	&0.635	&7.342	&19&S\cr
NGC 5347	&12.70	&2370	&SBab		&1.848	&1.525	&0.825	&0.508	&1.424	&11&S\cr
NGC 5506	&14.38	&1830	&Sa pec		&3.586	&9.369	&2.613	&0.267	&8.409	&14&M\cr
NGC 5643	&13.60	&990	&SABc		&1.844	&1.895	&1.028	&0.152	&19.490	&14&M\cr
NGC 5728	&13.40	&2790	&SABa		&2.453	&5.257	&2.143	&0.156	&8.163	&12&M\cr
NGC 6300	&13.08	&900	&SBb		&2.491	&---	&---	&0.279	&14.650	&2&L\cr
NGC 6890	&14.02	&2430	&SAb		&1.081	&---	&---	&0.107	&3.855	&1&L\cr
NGC 7130	&13.87	&4830	&Sa pec		&2.442	&---	&---	&0.188	&16.480	&1&L\cr
NGC 7314	&13.11	&1500	&SABbc		&2.401	&2.290	&0.954	&0.139	&3.736	&14&M\cr
NGC 7582	&13.57	&1560	&SBab		&1.745	&6.584	&3.772	&0.258	&49.100	&27&M\cr
NGC 7743	&13.28	&2040	&SB0/a		&1.871	&---	&---	&0.595	&0.791	&2&L\cr
Mrk 1		&14.96	&4800	&SB0/a		&1.491	&3.286	&2.204	&0.137	&2.531	&7&M\cr
Mrk 3		&13.34	&4110	&E2 pec		&2.351	&---	&---	&0.164	&3.770	&7&M\cr
Mrk 34		&14.76	&15450	&S		&3.000	&---	&---	&0.220	&0.809	&7&M\cr
Mrk 78		&15.00	&11145	&E/S0		&2.383	&---	&---	&0.196	&1.110	&7&M\cr
Mrk 176		&14.61	&8070	&S0/a pc	&1.111	&---	&---	&0.108	&0.694	&7&M\cr
Mrk 198		&14.73	&7170	&SAB0 pc	&4.370	&---	&---	&0.351	&0.624	&7&M\cr
Mrk 268		&14.66	&12300	&Sa		&4.410	&---	&---	&0.535	&1.381	&7&M\cr
Mrk 270		&14.05	&3090	&SAB0		&3.943	&---	&---	&0.541	&---	&7&M\cr
Mrk 348		&13.90	&4410	&SA0/a		&2.480	&---	&---	&0.247	&1.290	&7&M\cr
Mrk 423		&14.29	&9720	&S0?		&3.591	&---	&---	&0.790	&1.423	&25&S\cr
Mrk 463e	&14.22	&15150	&S pec		&3.231	&---	&---	&0.210	&2.184	&18&S\cr
Mrk 516		&16.50	&8519	&		&3.765	&---	&---	&0.640	&1.325	&25&S\cr
Mrk 573		&14.07	&5130	&SAB0		&2.089	&---	&---	&0.167	&1.088	&7&M\cr
Mrk 609		&14.12	&10260	&E/S0		&1.545	&3.333	&2.157	&0.170	&2.550	&25&S\cr
Mrk 612		&15.50	&6206	&SB0/a?		&1.864	&---	&---	&0.110	&1.159	&18&S\cr
Mrk 622		&14.40	&6840	&S0 pec		&7.656	&---	&---	&0.490	&1.281	&18&S\cr
Mrk 1066	&14.01	&3540	&SB0		&4.211	&---	&---	&0.320	&10.98	&21&S\cr
Mrk 1193	&16.50	&9600	&		&3.355	&---	&---	&0.190	&0.641	&13&L\cr
Mrk 1388	&15.70	&6390	&		&0.473	&---	&---	&0.048	&0.174	&8&S\cr
IC 1515		&14.80	&6870	&SBb		&1.524	&0.976	&0.641	&0.170	&0.566	&12&M\cr
IC 5063		&13.60	&3300	&SA0a pc	&3.656	&---	&---	&0.223	&5.337	&1&L\cr
MCG-5-23-16	&13.69	&2280	&S0		&1.533	&1.769	&1.154	&0.110	&---	&23&S\cr
UM 16		&17.00	&17400	&		&1.605	&2.653	&1.653	&0.130	&---	&18&S\cr
UM 82		&17.88	&15300	&		&1.802	&---	&---	&0.095	&---	&13&L\cr
UM 85		&17.29	&12300	&		&3.293	&---	&---	&0.174	&---	&13&L\cr
UM 103		&17.00	&13500	&		&3.113	&---	&---	&0.280	&---	&13&L\cr
UM 293		&16.40	&16800	&		&2.804	&---	&---	&0.239	&---	&13&L\cr
Fairall 4	&15.10	&15900	&Sb		&1.579	&---	&---	&0.101	&---	&13&L\cr
ESO 138-G1	&14.31	&2730	&E/S0		&2.219	&1.426	&0.643	&0.193	&---	&30&M\cr
ESO 103-G35	&14.53	&4050	&S0?		&5.651	&---	&---	&0.365	&2.314	&14&M\cr
TOL0514-415	&	&14700	&		&4.154	&---	&---	&0.300	&---	&13&L\cr
TOL0544-395	&14.90	&7500	&S0		&3.584	&---	&---	&0.239	&0.917	&13&L\cr
TOL0611-375	&	&11400	&		&4.115	&---	&---	&0.213	&---	&13&L\cr
WAS 49a		&	&18900	&		&1.923	&5.263	&2.737	&0.101	&0.438	&10&S\cr
POX 52		&17.20	&6420	&		&3.509	&8.000	&2.280	&0.288	&---	&24&M\cr
IZw 92		&15.20	&11340	&		&2.062	&3.448	&1.672	&0.200	&1.313	&18&S\cr
IIIZw 55	&15.40	&7507	&		&1.928	&---	&---	&0.172	&0.867	&7&M\cr
VZw 317		&15.77	&10200	&		&1.571	&---	&---	&0.660	&---	&25&S\cr
\tablenotetext{}{Morphological types were obtained from de Vaucouleurs et al. 1991
and Mulchaey 1994. The references correspond to the same namber as in Table 1.}
\enddata
\end{deluxetable}

\newpage

\begin{deluxetable}{lrrlrrrr}
\tablewidth{0pc}
\tablecaption{Comparison between Seyfert 1's and Seyfert 2's}
\tablehead{\colhead{Ratio}&\colhead{Sy1}&\colhead{Mean}&\colhead{STD}&
\colhead{Sy 2}&\colhead{Mean}&\colhead{STD}&\colhead{P(\%)}}
\startdata
60$\mu$m&       40&  43.61& 0.48&   52&   43.71& 0.60&  43.74\cr
Morph Types    &37&   --- &  ---&   52&     ---&  ---&  99.86\cr
[OII]/[NeIII]&  52&   1.73& 0.81&   68&    2.91& 1.37&  0.02\cr
[OII]/[OIII]&   52&   0.19& 0.11&   68&    0.26& 0.16&  22.59\cr
[OII]/[NeV]&    31&   1.54&  1.6&   21&    3.49& 2.35&  0.16\cr
[NeIII]/[NeV]&  31&   1.05& 0.74&   21&    1.55& 0.85&  10.87\cr
\tablenotetext{}{Column 1 gives the quantity that is being analysed,
column 2(5), 3(6) and 4(7) give the number of Seyfert 1's (Seyfert 2's)
with the measurement
available, their average value and standard deviation, respectively.
Column 8 gives the KS test significance for rejection of the hypothesis that the two
distributions are drawn from the same population.}
\enddata
\end{deluxetable}